%% file: main.tex
\documentclass[conference,a4paper]{IEEEtran}
\IEEEoverridecommandlockouts
\usepackage{cite}
\usepackage{amsmath,amssymb}
\usepackage{graphicx}
\usepackage{textcomp}
\usepackage{xcolor}
\definecolor{Locked}{RGB}{38,110,72}
\definecolor{Exposed}{RGB}{176,58,46}
\definecolor{Ink}{RGB}{70,70,70}
\usepackage{booktabs}
\usepackage{tabularx}
\newcolumntype{L}{>{\raggedright\arraybackslash}X}
\usepackage{url}
\usepackage{tikz}
\usepackage{pgfplots}
\pgfplotsset{compat=1.18}
\usetikzlibrary{arrows.meta,positioning,fit,backgrounds,calc}
\usepackage[hidelinks]{hyperref}
\usepackage{orcidlink}

\usepackage[acronym,nomain]{glossaries}
\setacronymstyle{long-short}
\glsdisablehyper
\loadglsentries{acronyms}

\begin{document}
\bstctlcite{IEEEexample:BSTcontrol}

\title{PULSAR-V: I/O-Aware Post-Update \\ Attestation for RISC-V}

\author{
\IEEEauthorblockN{
Niccolò Lentini\textsuperscript{1}\orcidlink{0009-0008-6639-7414},
Stefano Di Carlo\textsuperscript{1}\orcidlink{0000-0002-7512-5356},
Alessandro Savino\textsuperscript{1}\orcidlink{0000-0003-0529-7950}
}

\vspace{0.3em}

\IEEEauthorblockA{\textsuperscript{1}Control and Computer Engineering Department, Politecnico di Torino, Turin, Italy}

\IEEEauthorblockA{
niccolo.lentini@polito.it, stefano.dicarlo@polito.it, alessandro.savino@polito.it
}
}

\maketitle

\begin{abstract}
Post-update attestation proves that a device installed the intended firmware, not that the image stayed protected until the report was signed. In a multi-master \gls{soc}, a master outside the attested processor can corrupt update-critical memory. In contrast, the firmware identity, lifecycle state, and the \gls{cpu}'s protection evidence remain byte-identical to a healthy device's. We present PULSAR-V, a post-update attestation architecture which binds the authenticated update commit to the effective permissions and immutability of \emph{both} gates (\gls{pmp}/\gls{epmp} on the \gls{cpu}, the \gls{io} policy mediating \gls{dma}), sampled at the commit and at attestation, the \gls{io} gate latching its own commit-time state in hardware. In a RISC-V \gls{soc} prototyped on a ZCU102 \gls{fpga} (CV32E40S core with 16 \gls{pmp} regions, \gls{iopmp}-style unit, \gls{axi} write guard), a 256-byte write from the board's Arm \gls{dma} engine bypasses \gls{pmp} and corrupts the committed image: every conventional claim stays true, and PULSAR-V rejects the device on its \gls{io}-policy evidence alone. It also rejects an \emph{intact} device whose policy left the image writable, a verdict no integrity check can reach. The gate and its commit latch cost 436 \glspl{lut}, 4.6\,\% of the \gls{soc}, and evidence collection takes \(43.1\)\,k cycles independent of image size, \(41\times\) less than re-hashing \(16\,\mathrm{KiB}\).
\end{abstract}


\glsresetall

\section{Introduction}
\label{sec:intro}

Firmware updates on low-end devices follow a fixed sequence: the new image is staged, verified, booted, self-tested, and finally \emph{committed}, the instant the device declares it authoritative~\cite{rfc9019,mcuboot}. Post-update \textit{attestation} then proves the outcome to a remote backend, with a signed report binding the committed image digest to the lifecycle state and a rollback counter~\cite{rfc9334,assured,pure}. On a constrained core, the digest is \emph{recorded} at the commit $T_c$ and is merely reported when the device answers a backend challenge at $T_a$, because recomputing it takes time that is linear in the image size, which is not affordable for every attestation. The report thus describes the image as it was at $T_c$. Whether it stayed so across $W=[T_c,T_a]$ depends on the memory isolation in force during $W$, which the report covers only in part.

On RISC-V, isolation is implemented through \gls{pmp} and its \gls{epmp}/Smepmp extensions~\cite{riscvpriv}, configured in the core's control registers and checked during fetch, load, and store operations. Any other master accesses never reach those checks. Stopping them needs a unit on the \gls{io} path, such as the still-unratified \gls{iopmp}~\cite{iopmpspec}. Connected devices require such a unit by construction: the radio controller, upon receiving an \gls{ota} image, writes it to memory via \gls{dma}. A commodity Wi-Fi \gls{mcu} pairs a \gls{pmp} with several \gls{dma} channels and no \gls{io}-side permission check~\cite{gd32vw}, and such radio firmware is a demonstrated remote attack surface: Project Zero compromised a Broadcom controller over the air and reached host memory by \gls{dma}~\cite{overtheair}. A compromised controller that writes the active slot inside $W$ leaves every attested claim untouched, since the digest predates the write, the lifecycle state is unchanged, and the \gls{pmp} configuration is correct and locked. The report is truthful, and the image is corrupted.

Existing remedies do not close this gap. Re-hashing at $T_a$ detects damage at an $O(n)$ cost, but accepts a device that is writable but not yet written, and secure boot re-verifies the image only at the next boot. Reporting an \gls{iommu} enabled~\cite{healthattestation} says nothing about what it permits, and sampling its policy only at $T_a$ misses a relax-and-restore during $W$. Even a set lock bit is not immutable: with the \gls{rlb} bit of Smepmp set, a locked \gls{pmp} entry remains writable. To exclude modification across $W$, the verifier needs the permissions of \emph{every} gate mediating a bus master, at both ends of $W$, and proof that they could not change in between.

We present PULSAR-V, a post-update attestation architecture for low-end RISC-V systems that supplies this evidence. It attests \emph{exposure}: whether any bus master other than the attested core could have written the committed image during $W$. For both gates, \gls{pmp}/\gls{epmp} on the \gls{cpu} and the \gls{iopu} mediating \gls{dma}, the report carries the configuration at $T_c$, latched by the gate itself on the commit strobe, and at $T_a$, together with a reset count at both instants. The verifier appraises effective permissions and immutability rather than an enable or lock bit, and obliges the device to re-hash only the regions its own policy left writable. A device whose image is still intact is therefore rejected when nothing prevented it from being written. The threat is architecture-independent; the choice of RISC-V fixes the protection semantics to be appraised.
Our contributions are:
\begin{itemize}
    \item an attack in which a \gls{dma} master corrupts a committed image while identity, lifecycle, and locked-\gls{pmp} evidence stay byte-identical to a healthy device's, and the conditions R1--R5 under which it applies, checked against vendor documentation (Sec.~\ref{sec:threat});
   \item a verifier evidence composition that is necessary and sufficient to establish the absence of exposure across $W$, together with a hardware mechanism that captures commit-time policy state and uses the attested policy to determine when re-hashing is required (Secs.~\ref{sec:c2}--\ref{sec:mech});
    \item a prototype on a Xilinx ZCU102 that rejects an exposed but intact device, at 436 \glspl{lut} and a constant 43.1\,k cycles of evidence collection (Secs.~\ref{sec:impl}--\ref{sec:eval}).
\end{itemize}

\section{Related Work}
\label{sec:related}

No prior post-update attestation system, to our knowledge, binds the live \gls{cpu} \emph{and} \gls{io} isolation state protecting the committed image to the update commit. Prior work covers these elements separately.

\textbf{Update attestation.} SUIT~\cite{rfc9019} specifies update manifests, RATS~\cite{rfc9334} attester and verifier roles, and MCUboot~\cite{mcuboot} A/B-slot updates. ASSURED~\cite{assured} and PURE~\cite{pure} prove installation, reset, and erasure. All attest outcomes or memory contents, not the state protecting the image. RASUES~\cite{rasues}, closest in intent, attests across updates with Linux \gls{ima} and a \gls{tpm}, but its software adversary excludes isolation configuration and non-\gls{cpu} masters.

\textbf{Runtime and periodic attestation.} Self-measurement~\cite{erasmus,seed} re-hashes memory on a schedule, at $O(n)$ per measurement. Hybrid attesters monitor the bus, \gls{dma} included: RATA~\cite{rata} records that a region \emph{was} modified, VRASED~\cite{vrased} and IDA~\cite{ida} fault external writes to protected regions, and CASU~\cite{casu} prevents them. Their monitors hardwire regions and policy at design time, leaving no configuration to attest to, whereas PULSAR-V targets gates whose policy is set at deployment, is reconfigurable, or is split across units. Their verdicts are also triggered by damage, not exposure, so a writable but unwritten device passes. DMA'n'Play~\cite{dmanplay} uses a \gls{dma} engine to \emph{perform} attestation, not as the adversary.

\textbf{Isolation-state attestation.} EPA-RIMM~\cite{eparimm} and ROMP~\cite{romp}, a patent application, attest live protection registers and detect changes while evidence is valid, but only on the \gls{cpu} side and unbound to an update.

\textbf{\gls{io} configuration evidence.} Windows Health Attestation~\cite{healthattestation} reports whether the \gls{iommu} is enabled, not what it permits, and shipped x86 firmware shows the two diverge: in four vulnerabilities disclosed in 2025, firmware asserts \gls{dma} protection it never configures~\cite{dmafail}, and a locked interface reported disabled pre-boot protection as enabled~\cite{lockedbios}. Lifecycles in the style of \gls{tdisp}~\cite{amdtdisp} hash the \emph{device}'s own configuration, not the mediation unit. Arm \gls{cca}~\cite{armcca} keeps device permissions in its \gls{smmu}, but its platform claim lists ``chosen implementation options''~\cite{ccatoken}, not live tables. An enabled \gls{iommu} is not a validated one~\cite{thunderclap}.

\textbf{Multi-master isolation.} On Armv8-M, TrustZone already gates bus masters through master security and memory protection controllers~\cite{sie200}, yet Arm's PSA attestation token carries lifecycle state and software measurements, not controller configuration~\cite{psatoken}. On RISC-V, most enclave frameworks, such as Keystone~\cite{keystone}, isolate through PMP alone, and a recent survey treats \gls{dma} protection as complementary~\cite{rvteesurvey}. \gls{iopmp} and its implementations~\cite{iopmpspec,iopmpng,iopmpyuan,siopmp}, CURE's bus filters~\cite{cure}, HECTOR-V's \gls{io} paths~\cite{hectorv}, WorldGuard~\cite{worldguard} and vendor units~\cite{esp32c3trm} fill that role, though the \gls{mcu} initiator side is argued to remain under-protected~\cite{securewheels}. All of these \emph{enforce} a policy, and none reports it to a verifier. The survey~\cite{rvteesurvey} lists attestation only as a cross-architecture concern. It credits WorldGuard with eliminating configuration-driven vulnerabilities, yet its checkers are software-locked or dynamic, and the lock holds only until a reset~\cite{worldguard}: precisely the state PULSAR-V attests to. Lacking a common register interface, PULSAR-V maps each platform's enforcing registers and permission semantics onto a common attestation contract (Sec.~\ref{sec:mech}).

\section{Threat Model and Attack}
\label{sec:threat}

\begin{figure}[tb]
    \centering
    \includegraphics[width=0.99\linewidth]{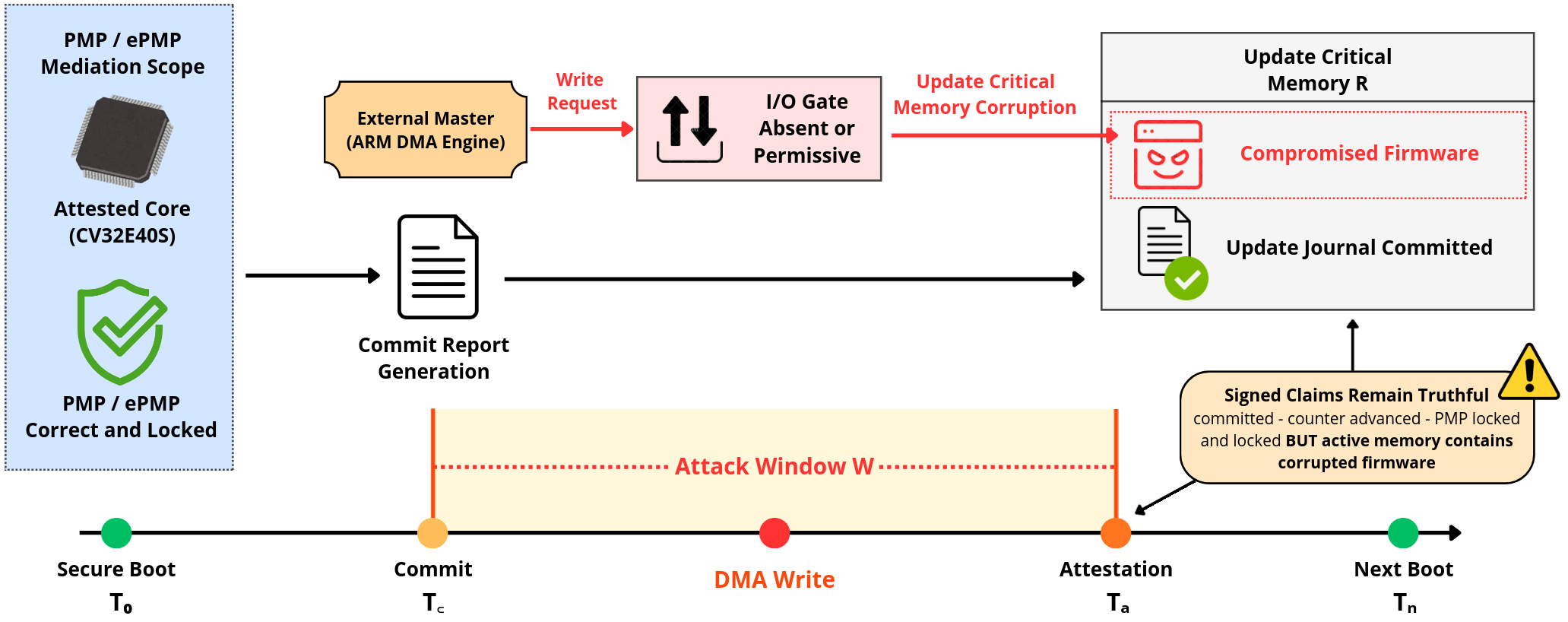}
    \caption{Attack model and vulnerable interval. \acrshort{pmp}/\acrshort{epmp} mediates only the attested core. The external \acrshort{dma} master reaches update-critical memory through a separate \acrshort{io} path. A write during $W=[T_c,T_a]$ corrupts the active firmware while the commit record and \acrshort{cpu}-side evidence stay unchanged.}
    \label{fig:attack-model}
\end{figure}

We consider a RISC-V \gls{soc} that has completed an authenticated update and is attested by a remote backend (Fig.~\ref{fig:attack-model}). Besides the \emph{attested core}, which runs update and attestation firmware, it has an \emph{external master} (a \gls{dma} engine, communication controller, or second processor) whose transactions bypass that core. $R$ denotes the update-critical regions, $T_c$ the commit time, $T_a$ the instant the attester samples its evidence for a verifier challenge, and $W=[T_c,T_a]$. The backend appraises the signed samples afterward. The adversary can program the external master, compromise the attested core's firmware (on a RISC-V part with only \gls{mmode}, it is everything outside the \gls{rom} anchor), and reprogram any unlocked isolation configuration. It \textbf{cannot} modify \gls{rom}, break cryptography, or bypass a correctly configured and locked protection unit. As in SMART~\cite{smart}, the attestation routine and key reside in immutable memory, with the key accessible only during attestation. The adversary, therefore, cannot forge reports and must keep them truthful while the image diverges.
The window opens at the commit: at $T_c$ the device measures the active image and records the lifecycle transition. Secure boot and the signature check ran earlier, at activation. Inside $W$, the external master writes the active slot, unseen by \gls{pmp}, and the write lands unless a gate denies it. This requires neither physical access nor code on the attested core, which remains honest but unaware of the transaction, so its truthful report conceals the compromise. Such masters sit outside the core's control by architecture~\cite{iopmpspec}.

The attack works wherever \textbf{R1:} at least two bus masters can write memory; \textbf{R2:} some region in $R$ is writable by a master other than the attested core; \textbf{R3:} no \gls{io}-side mediation exists, \emph{or} its policy is reconfigurable after boot and unlocked; \textbf{R4:} the report carries a measurement \emph{recorded} at or before $T_c$, as in any $O(1)$ attester; and \textbf{R5:} the interval to the next re-verification is non-zero. Under R1--R5, a corrupted device is byte-identical to a healthy one for every evidence profile lacking \gls{io} policy state. A countermeasure must break R3 by mediating every master, and lift the indistinguishability left by R4 by attesting that the policy is bound to the commit.

R2 and R3 hold in deployed parts for three reasons, none of them a misconfiguration that testing would catch. The updated transport is a \gls{dma} master with write access to the update region, which a compromise inherits (R2). Low-end \glspl{mcu} ``lack sophisticated features, such as specialized memory management or protection features''~\cite{smart}, so downloadable modules, scripting runtimes, and \gls{ota}-delivered configuration that run in \gls{mmode} can reprogram the gate. Deployed \glspl{mpu} leave peripheral masters unrestricted, and where a gate exists, as the NXP \texttt{sysMPU}, it ``causes firmware hangs'' so often that developers disable it~\cite{mpuwild}. The mass-produced GD32VW553, shipped with a secure-boot and firmware-upgrade application note, arbitrates ``the competition between these two masters'' on its \gls{ahb} interfaces, and its 614 pages describe no \gls{io}-side permission check~\cite{gd32vw}. The ESP32-C3 enables \emph{and} locks its Permission Control unit by default~\cite{esp32c3trm}, yet still needs evidence: both are build-time options applied by startup code, not \gls{rom} invariants, and the lock does not survive reset, so a fleet mixes isolation states behind identical firmware evidence. The first part needs the gate, the second the evidence.

Prevention is the natural alternative. Where \gls{rom} locks the gate irreversibly, R3 fails, and CASU-style prevention~\cite{casu} is the answer. It requires three conditions at once: a gate to lock, a lock that is a \gls{rom} invariant rather than a value startup code writes, and a lock that outlives a reset. The GD32VW553 lacks the first; the ESP32-C3, the other two, as enable and lock are build-time options applied by the startup code, and neither survives reset~\cite{esp32c3trm}. One firmware image can thus ship a fleet whose isolation states differ, while all firmware measurements agree. Prevention says whether a device \emph{can} be protected, attestation whether this one \emph{was}.

\section{Evidence Composition: Necessity and Sufficiency}
\label{sec:c2}

$M$ is the set of bus masters that can write memory and $x@t$ the value of $x$ sampled at instant $t$. Each $m\in M$ is mediated by a \emph{gate} $G(m)$ (here $G(\mathit{cpu})=\mathrm{PMP}$ and $G(\mathit{ext})=\mathrm{IOPU}$) with configuration $\mathit{cfg}(G)$, all of its permission registers, and reset count $\mathit{rst}(G)$. We assume every access by $m$ passes $G(m)$ and that its outcome depends only on $\mathit{cfg}(G)$ (Secs.~\ref{sec:mech} and~\ref{sec:limits}). $\mathit{imm}(c)$ is the platform's immutability predicate over a configuration $c$. On the \gls{cpu} gate, it is $\mathit{lock}\wedge\neg\mathrm{RLB}$, where $lock$ means the entries covering $R$ and every lower-numbered one have the configuration bit \texttt{L} set (since an unlocked entry of higher priority could reopen $R$), and \texttt{mseccfg.RLB}, which is the rule-lock bypass that allows M-mode writes, is clear. On an \gls{io} gate, the lock must also cover the gate's own enable bit (see Sec.~\ref{sec:mech}). The verifier must establish \textbf{P}: \emph{throughout $W$, the configuration in force at every gate $G(m)$ denied every unauthorized master $m\neq\mathit{cpu}$ write access to every $r\in R$}. Unlike RATA's ``no write occurred'' or CASU's ``no write can occur'', P is the absence of unmediated exposure across $W$, decided from policy state rather than the image.

For each gate, the device reports
\[ E(G)=\bigl\langle \underbrace{\mathit{cfg}(G)@T_c,\,\mathit{cfg}(G)@T_a}_{\text{policy snapshots}},\; \underbrace{\mathit{rst}(G)@T_c,\,\mathit{rst}(G)@T_a}_{\text{reset counts}} \bigr\rangle \]
together with $E(\mathit{fw})$, the firmware and lifecycle evidence fixing \emph{which} image the verdict concerns. Immutability needs no sample of its own, since it is computed from $\mathit{cfg}(G)$, while the reset count lies outside it and is sampled twice. On the \gls{io} gate, the commit-time state is latched in a dedicated bank and read back by the attester only at $T_a$ (see Sec.~\ref{sec:mech}). Thus, freshness and continuity do not rely on firmware honesty. The \gls{cpu} gate, which P does not rely on, has no such hardware bank.

This evidence is necessary: without $E(G(m))$, P is undecidable for any attester satisfying R4. Consider two executions $A$ and $B$ differing only in $G(m)$: in $A$ the configuration denies $m$ throughout $W$ and no write occurs, in $B$ it permits $m$, which writes some $r\in R$ inside $W$. $E(\mathit{fw})$, the lifecycle claims and the other gates' evidence, all fixed at $T_c$ before the write, are identical in both. A verifier without $E(G(m))$ cannot separate them. R4 limits scope: re-hashing at $T_a$ separates $A$ from $B$, at the $O(n)$ cost R4 avoids. 

\begin{figure}[htb]
\centering
\begin{tikzpicture}[x=1cm, y=1cm, font=\scriptsize, inner sep=1pt]
  
  \def\xa{1.42}\def\xb{5.62}      % window: commit .. attestation
  \def\ch{0.185}                  % bar half-height
  \def\xv{7.95}                   % verdict column
  \def\rowa{0}\def\rowb{-0.5}\def\rowc{-1.0}\def\rowd{-1.5}\def\rowe{-2.0}

  \def\ca{6.05}\def\cb{6.5}\def\cc{6.95}\def\cd{7.4}

  \newcommand{\wmark}[2]{\fill[Ink] (#1-0.075,#2+0.125) -- (#1+0.075,#2+0.125) -- (#1,#2+0.02) -- cycle;}
  \newcommand{\lane}[2]{\node[anchor=east] at (\xa-0.1,#1) {#2};}
  \newcommand{\seg}[4]{\fill[#4, rounded corners=0.8pt]
    ({\xa+#2*(\xb-\xa)},{#1-\ch}) rectangle ({\xa+#3*(\xb-\xa)},{#1+\ch});}
  \newcommand{\band}[1]{\fill[Ink!5] (-0.02,{#1-0.24}) rectangle (8.62,{#1+0.24});}
  \newcommand{\yes}[2]{\node[Locked] at (#1,#2) {\checkmark};}
  \newcommand{\no}[2]{\fill[Exposed!15, rounded corners=1pt]
      (#1-0.19,#2-0.19) rectangle (#1+0.19,#2+0.19);
    \node[Exposed, font=\bfseries] at (#1,#2) {$\times$};}
  \newcommand{\acc}[1]{\node[anchor=west, Locked, font=\tiny\bfseries] at (\xv,#1) {accept};}
  \newcommand{\rej}[1]{\node[anchor=west, Exposed, font=\tiny\bfseries] at (\xv,#1) {reject};}

  \band{\rowb}\band{\rowd}

  \foreach \cx/\ct in {\ca/match (1), \cb/freshness (2), \cc/immutability (3), \cd/continuity (4)}
    {\node[rotate=40, anchor=west, font=\tiny] at (\cx-0.06,0.33) {\ct};}
  \node[rotate=40, anchor=west, font=\tiny\itshape] at (\xv-0.06,0.33) {verdict};
  \draw[Ink!45, line width=0.4pt] (\ca-0.28,0.27) -- (8.62,0.27);

  \draw[dashed, Ink!55] (\xa,0.30) -- (\xa,-2.2);
  \draw[dashed, Ink!55] (\xb,0.30) -- (\xb,-2.2);
  \node[anchor=south] at (\xa,0.36) {$T_c$};
  \node[anchor=south] at (\xb,0.36) {$T_a$};
  \draw[<->, Ink, line width=0.4pt] (\xa,0.16) -- node[above, font=\tiny, inner sep=0.5pt] {$W$} (\xb,0.16);

  \lane{\rowa}{defended}
  \seg{\rowa}{0}{1}{Locked}
  \yes{\ca}{\rowa}\yes{\cb}{\rowa}\yes{\cc}{\rowa}\yes{\cd}{\rowa}
  \acc{\rowa}

  \lane{\rowb}{\textcircled{\tiny 1} open}
  \seg{\rowb}{0}{1}{Exposed!30, draw=Exposed}
  \wmark{4.3}{\rowb}
  \no{\ca}{\rowb}\yes{\cb}{\rowb}\yes{\cc}{\rowb}\yes{\cd}{\rowb}
  \rej{\rowb}

  \lane{\rowc}{\textcircled{\tiny 2} tidied}
  \seg{\rowc}{0}{0.33}{Exposed!30, draw=Exposed}
  \seg{\rowc}{0.33}{1}{Locked}
  \wmark{2.1}{\rowc}
  \yes{\ca}{\rowc}\no{\cb}{\rowc}\yes{\cc}{\rowc}\yes{\cd}{\rowc}
  \rej{\rowc}

  \lane{\rowd}{\textcircled{\tiny 3} restored}
  \seg{\rowd}{0}{0.2}{Locked}
  \seg{\rowd}{0.2}{0.5}{Exposed!30, draw=Exposed}
  \seg{\rowd}{0.5}{1}{Locked}
  \wmark{2.9}{\rowd}
  \yes{\ca}{\rowd}\yes{\cb}{\rowd}\no{\cc}{\rowd}\yes{\cd}{\rowd}
  \rej{\rowd}

  \lane{\rowe}{\textcircled{\tiny 4} reset}
  \seg{\rowe}{0}{0.16}{Locked}
  \seg{\rowe}{0.16}{0.5}{Exposed!30, draw=Exposed}
  \seg{\rowe}{0.5}{1}{Locked}
  \wmark{2.75}{\rowe}
  \draw[Ink, line width=0.5pt] ({\xa+0.16*(\xb-\xa)},-2.21) -- ({\xa+0.16*(\xb-\xa)},-1.79);
  \node[anchor=north, font=\tiny] at ({\xa+0.16*(\xb-\xa)},-2.2) {reset};
  \yes{\ca}{\rowe}\yes{\cb}{\rowe}\yes{\cc}{\rowe}\no{\cd}{\rowe}
  \rej{\rowe}

  \node[anchor=north west, font=\tiny] at (-0.02,-2.52) {%
    \tikz[baseline=-0.55ex]{\fill[Locked] (0,-0.07) rectangle (0.26,0.07);}~denied\quad
    \tikz[baseline=-0.55ex]{\fill[Exposed!30] (0,-0.07) rectangle (0.26,0.07);
      \draw[Exposed, line width=0.4pt] (0,-0.07) rectangle (0.26,0.07);}~permitted\quad
    $\blacktriangledown$~external write lands\quad
    $\times$~the check that rejects};
\end{tikzpicture}
\caption{Each strategy passes every check to its left and is rejected by exactly one, so dropping that check admits it. \textcircled{\scriptsize 1} is
open throughout yet locked, so only \emph{match} sees it;
\textcircled{\scriptsize 2} is tidy in the report, so only the latched
$\mathit{cfg}@T_c$ does; \textcircled{\scriptsize 3} restores the gate but
never locks it; \textcircled{\scriptsize 4} clears the lock with a reset and
falls to the counter. \emph{Defended} and \textcircled{\scriptsize 1} are the
measured rows of Table~\ref{tab:staircase}.}
\label{fig:window}
\end{figure}
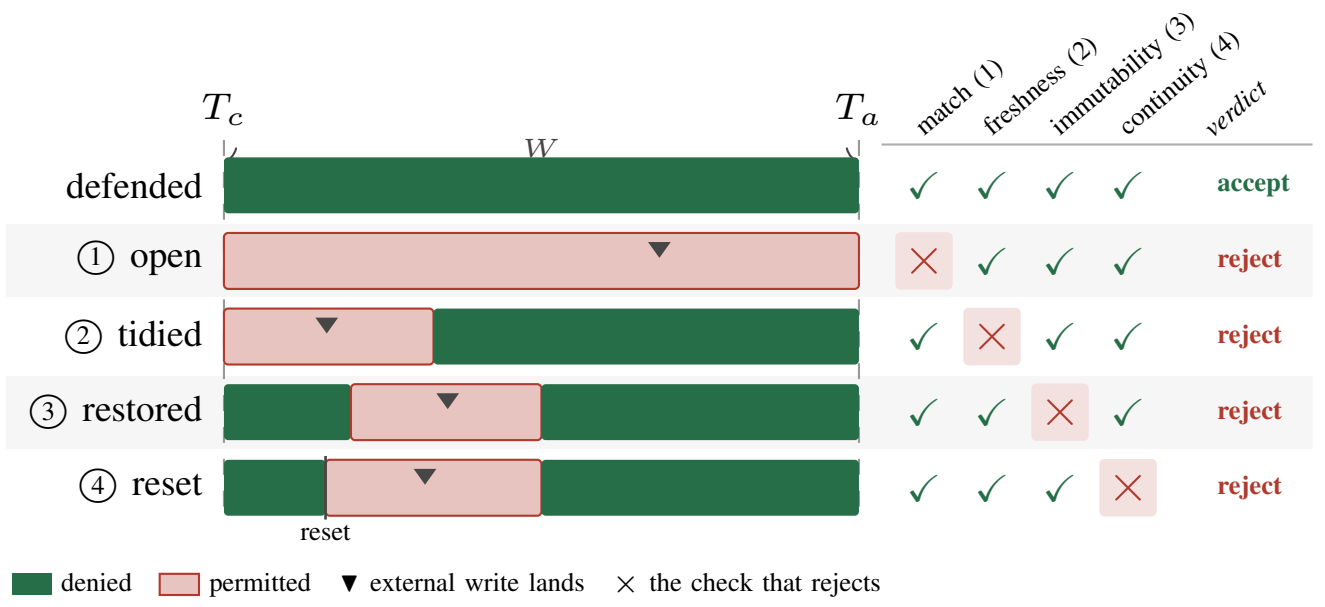

It is also sufficient. Suppose the verifier checks, for every gate $G=G(m)$: (1)~\emph{match}, $\mathit{cfg}(G)@T_a$ equals the golden configuration, which denies $m$ write access to all of $R$; (2)~\emph{freshness}, $\mathit{cfg}(G)@T_c=\mathit{cfg}(G)@T_a$; (3)~\emph{immutability}, $\mathit{imm}(\mathit{cfg}(G)@T_a)$; (4)~\emph{continuity}, $\mathit{rst}(G)@T_c=\mathit{rst}(G)@T_a$. Then P holds. Assume, for contradiction, that the configuration at some $t\in W$ permitted $m$ to write $r$; by (1), $\mathit{cfg}(G)@t\neq\mathit{cfg}(G)@T_a$, so $\mathit{cfg}(G)$ is not constant on $W$. But (2) and (3) make the gate immutable at both ends of $W$, as $\mathit{imm}$ reads only the configuration. An immutable gate changes only across a reset, and (4) excludes one, a contradiction. Check (4) turns ``immutable at both ends'' into ``immutable throughout'', since a reset clears PMP locks by specification~\cite{riscvpriv}, as it does the \gls{io} locks we examined. The counter sits outside the reset domain and is sound only if monotonic across all resets, including power loss. Where the reference declares an exposure legitimate, (1) is relaxed for that region, and a re-hash replaces P there with an integrity verdict.

The four checks are not redundant. Our \emph{match} compares the whole configuration, lock and enable bits included, so a locked golden configuration makes (1) subsume (3). We keep (3) because a reference pinning permissions only would admit a reconfigurable gate, and under such a reference, each check is needed (Fig.~\ref{fig:window}). Sec.~\ref{sec:eval} exercises all four.

\section{Mechanism}
\label{sec:mech}

Besides the firmware state, the report includes each gate's configuration at the commit and at attestation, as raw register words and their digests, plus the reset count at both instants. At the commit, the device locks the \gls{io} gate, samples the \gls{cpu} gate, and hashes the slot before strobing the latch, so the latched configuration is the one in force while the slot was hashed. The \gls{io} gate commit-time values are not held by the attester. On the strobe, the gate latches its configuration and reset count into a read-only bank, which is read at attestation. The latch is one-shot until reset, so a master reopening the gate inside $W$ cannot re-latch to conceal it, and a reset that clears the latch is witnessed by the counter, which shares the core's reset line. Both ends of each \gls{io}-gate comparison are therefore hardware states. The \gls{cpu} gate has no such bank. One signature over a fixed-order encoding covers every claim and the verifier's challenge, so snapshots cannot be mixed across sessions.

A digest shows that a configuration matches the enrolled one, not what it permits, so the verifier must decode the raw words. A locked critical region with \gls{rlb} set is not immutable, and the gap between an asserted lock and effective immutability appears on ratified and hand-written gates alike: an attester that reports only the lock bit misses the \gls{rlb} case, and an early revision of our own \gls{iopu} left \texttt{CTRL} writable under \texttt{LOCK} (\texttt{CTRL} holds the unit's enable bit and \texttt{LOCK} freezes the region table until reset). The gate could thus be disabled, reopened, and restored within $W$, leaving identical snapshots, asserted locks, and an unchanged reset count. For each update-critical range, the verifier determines from each sample whether the relevant master can write it and whether this permission can change without a reset, using \gls{pmp} semantics for \gls{mmode} or the unit's own matching rules.
The lowest-numbered matching entry decides. An entry binds \gls{mmode} only with \texttt{L} set, \texttt{mseccfg.MMWP} (the machine-mode whitelist policy) denies an unmatched access, and \texttt{RLB} defeats every state of \texttt{L}. \texttt{mseccfg.MML}, which changes the meaning of the permission bits, and the reserved \texttt{R}{=}0/\texttt{W}{=}1 encoding fall outside that model and are rejected rather than guessed at, as are the report flags that contradict the raw words. Any exposure is treated as a violation unless it is explicitly enrolled. Staging remains outside $W$, while an enrolled exposure still requires that snapshots remain unchanged and match the enrolled policy. A \gls{cpu} gate sample contains \texttt{pmpcfg0}--\texttt{3}, \texttt{pmpaddr0}--\texttt{15}, and \texttt{mseccfg} (84\,B), while for the \gls{io} gate it comprises control word, region count, and per-region base, \gls{napot} mask, and permissions (56\,B). These registers, their decoding, the immutability predicate, and the protected ranges form the attestation contract. Porting to an \gls{iopmp} or another gate requires redefining this contract, while the four checks of Sec.~\ref{sec:c2} remain unchanged. A gate with no commit bank still admits them, but its $\mathit{cfg}@T_c$ is software-reported, so P rests on the attester's honesty.

\begin{figure*}[tb]
  \centering
  \includegraphics[width=0.8\textwidth]{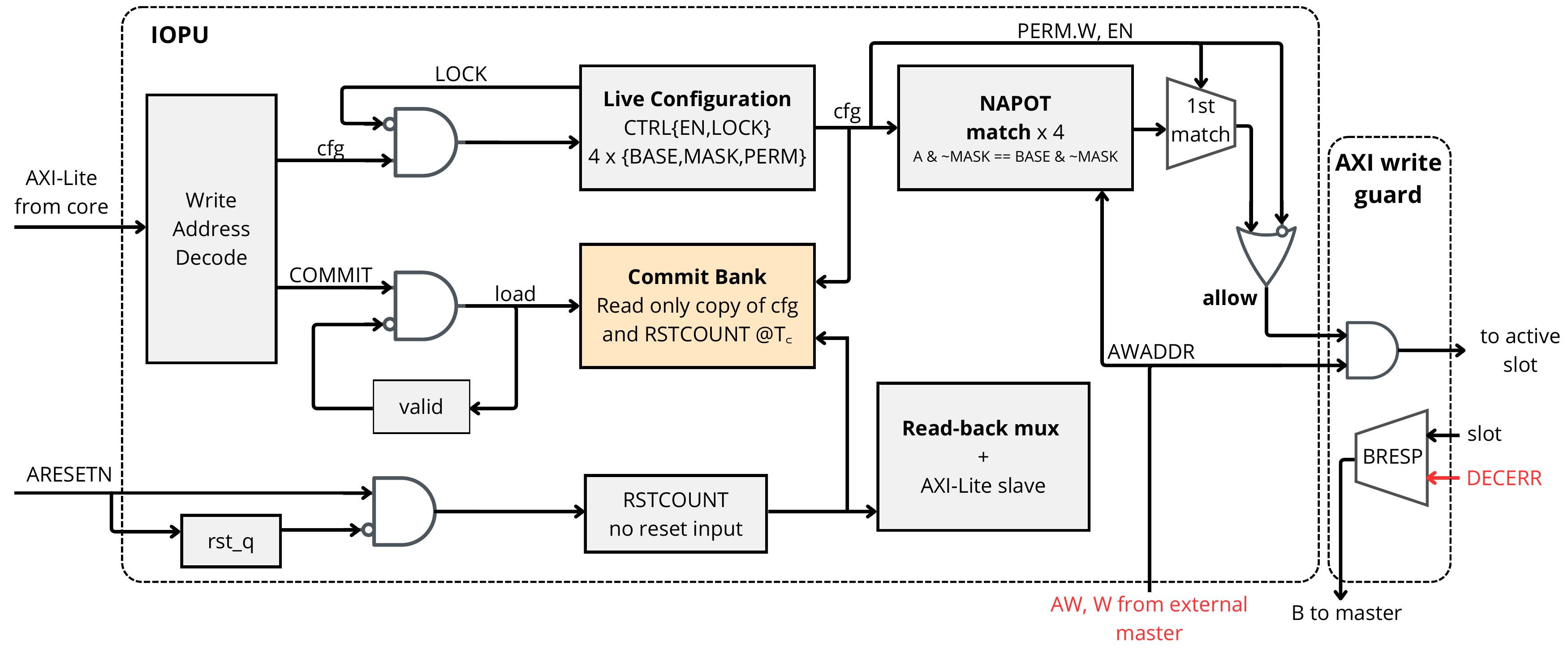}
  \caption{Hardware added by PULSAR-V. \texttt{LOCK} gates every write to the live configuration until reset. A \texttt{COMMIT} write, one-shot per reset, latches that configuration and \texttt{RSTCOUNT} into a read-only bank, so $\mathit{cfg}@T_c$ is hardware state. \texttt{RSTCOUNT}'s flip-flops have no reset input, so no reset rewinds it. The \acrshort{napot} comparators decide each external write, and the guard answers \texttt{DECERR} when it is denied.}
  \label{fig:arch}
\end{figure*}

The attested policy triggers re-hashing rather than replacing it. The device recomputes a region digest (which costs $O(n)$) only when its own \gls{io} policy left that region reachable by the external master, giving three cases: (i) the attested policy denies writes, so, given (1)--(4), the digest recorded at $T_c$ settles integrity and no re-hash is done; (ii) the region was exposed and the reference declares that exposure, so the re-hash must match the recorded digest; (iii) the exposure is undeclared and the verifier rejects on policy alone (the device re-hashes anyway, not knowing the reference). The verifier derives the same obligation from the decoded samples, so an attester that omits a required re-hash is rejected.
The comparison point is the commit, so $W$ starts at $T_c$ rather than at a previous attestation as in periodic schemes. The challenge signed with the report ties recency to the session, avoiding replay of snapshots taken before the policy was relaxed.
Appraising live policy is meaningful only if the appraised registers determine access. A testbench replays access decisions from the sampled configuration and compares them with the enforcement output across disabled, read-only, widened, and post-lock configurations. A second check verifies that the lock prevents the control-register escape.

\section{Implementation}
\label{sec:impl}

The evaluation platform is a Xilinx ZCU102, with a complete \gls{soc} synthesized into its programmable logic (Fig.~\ref{fig:arch}). The Arm \gls{ps} is not part of that \gls{soc} and supplies only the clock, reset, and the adversary. The attested device is a CV32E40S soft core (RV32IMC, firmware entirely in \gls{mmode}) at 25\,MHz with 16 \gls{pmp} regions and \texttt{mseccfg}, whose data bus reaches, through an address decoder and bridges from \gls{obi} to \gls{axi} Lite, the \gls{iopu} (4 \gls{napot} regions, lock bit), a UART console, unified instruction/data \gls{bram}, and a separate \gls{bram} holding the active slot. Firmware is preloaded into \gls{bram}, and there is no debug master, so the core's only external interface is its console.

The \gls{iopu} exposes a memory-mapped region table (enable, lock, per-region base, \gls{napot} mask, and permissions), the one-shot commit latch of Sec.~\ref{sec:mech}, and a combinational check port. Unmatched addresses are permitted (a production \gls{iopmp} is fail-closed), so only programmed regions are mediated. An \gls{axi}4 write guard on the external master's path checks each transaction's start address on that port, passing permitted writes and returning a decode error (\texttt{DECERR}) for denied ones. 

The adversary is the board's own \gls{dma} engine: channel~0 of the ZynqMP low-power-domain ADMA, programmed by the Arm \gls{ps} and reaching the fabric through \texttt{M\_AXI\_HPM0\_LPD}, the path the Arm cores' own stores take, so mediating it needed no \gls{rtl} change. A shim confines the incoming window to the 4\,KiB active slot, forwards writes through the guard, and returns \texttt{DECERR} unchanged. Each attack copies 256 bytes into the slot.

\section{Evaluation}
\label{sec:eval}

Table~\ref{tab:staircase} reports results from the routed design. In \emph{defended}, the gate blocks the \gls{dma} transfer, \emph{open} and \emph{toctou} corrupt the slot, and \emph{exposed} leaves it writable but issues no write. Starting from P0 (signature, device identity, and image digest), the cumulative profiles add slot and version (P1)~\cite{mcuboot}, lifecycle state (P2)~\cite{pure}, live \gls{pmp}/\gls{epmp} state (P3)~\cite{romp}, and update-bound \gls{io} policy (P4). P0--P3 accept every scenario, so they share one column.

\begin{table}[t]
\centering
\caption{Measured on hardware. Writes come from the board's \acrshort{dma} engine, not the attested core. Here \texttt{5a} is the staged image byte, \texttt{ef} the low byte of \texttt{0xDEADBEEF}. Only P4 discriminates.}
\label{tab:staircase}
\normalsize
\setlength{\tabcolsep}{4pt}
\begin{tabularx}{\columnwidth}{@{}lLccc@{}}
\toprule
Scenario & \acrshort{io} policy for the slot & \texttt{slot[0]} & P0--P3 & P4 \\
\midrule
defended & read-only, locked      & \texttt{5a}$\rightarrow$\texttt{5a} & \textsc{acc} & \textsc{acc} \\
\textbf{exposed} & \acrshort{dma}-writable, \textbf{no write} & \texttt{5a}$\rightarrow$\texttt{5a} & \textsc{acc} & \textbf{\textsc{rej}} \\
open     & \acrshort{dma}-writable           & \texttt{5a}$\rightarrow$\texttt{ef} & \textsc{acc} & \textbf{\textsc{rej}} \\
toctou   & reopened after commit  & \texttt{5a}$\rightarrow$\texttt{ef} & \textsc{acc} & \textbf{\textsc{rej}} \\
\bottomrule
\end{tabularx}
\end{table}

Each row of Table~\ref{tab:staircase} is one run per bitstream load with a fresh 128-bit challenge. Five campaigns (20 runs) gave identical verdicts and failing checks, and the raw console logs are included in the artifact. The defended device is accepted for every profile, so the added evidence does not rule out a correct device. Both corrupted devices pass every firmware-identity, lifecycle, and \gls{cpu}-isolation check and are rejected only on the update-bound \gls{io}-policy evidence. \emph{Open} fails \emph{match} (1); \emph{toctou}, unlocked and still open at $T_a$, also fails \emph{freshness} (2) against the gate's latched bank and \emph{immutability} (3). A variant that closes the gate again before $T_a$ (lane \textcircled{\scriptsize 2} of Fig.~\ref{fig:window}) would pass \emph{match} and fall to that latched digest alone.

Exposure alone suffices for rejection. In \emph{exposed}, the slot remains \gls{dma}-writable, but no transfer is issued, so the image stays byte-identical, and both \gls{iopu} snapshots agree, yet \emph{match} (1) alone rejects the device. Re-hashing, RATA, and secure boot accept it, with no detectable damage. The verdict rests on the attested policy before damage is done, and the binding of Sec.~\ref{sec:mech} rejects replayed reports under a fresh challenge.

A deployment that must leave a region reachable enrolls that exposure in its reference. Against such a reference, the \emph{exposed} and \emph{open} runs diverge on the obliged re-hash. The exposed-but-intact device is accepted, its recomputed digest matching the recorded one, and the written device is rejected on the same comparison. Declaring an exposure does not license drift: \emph{toctou} is still rejected. A defended fleet keeps the $O(1)$ path, paying 43\,093 cycles per attestation instead of 487\,607, $11.3\times$ less than unconditional re-hashing.
In area, the \gls{io}-side enforcement logic that blocks this attack costs 261 \glspl{lut} and 492 \glspl{ff}, while the commit latch costs a further 175 \glspl{lut} and 449 \glspl{ff}. Together they add 436 \glspl{lut}, 4.6\,\% of the 9\,398-\gls{lut} \gls{soc} (4\,439 \glspl{ff}, 18 \gls{bram}, 3 \gls{dsp} slices; 3.4\,\% of the XCZU9EG). Per region, the 4-region enforcement costs ${\approx}65$ \glspl{lut} against ${\approx}314$ for each of the 16 \gls{pmp} regions, ${\approx}4.8\times$ less, the latter being the 5\,016-\gls{lut} difference between core builds with 16 and 0 regions, divided by 16. The gates differ in function (the \gls{pmp} checks instruction and data accesses with \gls{tor}, \gls{napot}, and \gls{na4} matching, priority, and execute permission, while the \gls{io} gate has four \gls{napot} regions on one write port), so the ratio bounds the cost of adding \gls{io} mediation rather than comparing equal work. Timing closes with 8.44\,ns of slack over 8\,814 endpoints. The 25\,MHz clock is set by the \gls{ps} clock divider, the critical path lying in the stock core. The evidence added over a firmware-hash-only attester is 43.1\,k cycles per attestation (28\,601 for both \gls{io} banks, 14\,492 for the \gls{pmp} sample) plus 14\,491 at the commit, 19\,\% of signing with \gls{hmac} over \gls{sha256}, a share of the asymmetric signature it stands in for (Sec.~\ref{sec:limits}) would only reduce.

\begin{figure}[t]
\centering
\begin{tikzpicture}
\begin{loglogaxis}[
  width=\columnwidth, height=5.0cm,
  font=\small,
  xlabel={image size (bytes)},
  ylabel={attestation cost (cycles)},
  xlabel style={font=\small, yshift=1.0mm},
  ylabel style={font=\small, yshift=-1.0mm},
  tick label style={font=\small},
  xmin=700, xmax=2200000,
  ymin=15000, ymax=300000000,
  xtick={1024,4096,16384,262144,1048576},
  xticklabels={1\,KiB,4\,KiB,16\,KiB,256\,KiB,1\,MiB},
  ytick={1e5,1e6,1e7,1e8},
  grid=major, grid style={gray!22},
  legend style={at={(0.02,0.97)}, anchor=north west, font=\small,
                draw=gray!50, fill=white, row sep=-1pt, inner sep=2pt},
  axis lines=left,
]

\addplot[mark=*, mark size=1.5pt, thick, Exposed]
  coordinates {(1024,115880) (4096,443720) (16384,1755080)};
\addlegendentry{re-hash image, $O(n)$}
\addplot[mark=none, thick, Exposed, dashed, forget plot]
  coordinates {(16384,1755080) (1048576,112325120)};

\addplot[mark=none, very thick, Locked]
  coordinates {(700,43093) (2200000,43093)};
\addlegendentry{isolation evidence, $O(1)$}

\draw[<->, gray!70, thin] (axis cs:16384,43093) -- (axis cs:16384,1755080);
\node[font=\small, anchor=west, gray!35!black] at (axis cs:18000,230000)
  {$41\times$};
\draw[<->, gray!70, thin] (axis cs:1048576,43093) -- (axis cs:1048576,112325120);
\node[font=\small, anchor=east, gray!35!black] at (axis cs:1000000,1600000)
  {${\approx}2600\times$};
\end{loglogaxis}
\end{tikzpicture}
\caption{Cost of isolation evidence against re-hashing. Re-hashing is linear in image size. The isolation snapshots hash a fixed 196\,B of registers at attestation. Solid: measured on the board; dashed: extrapolated at ${\approx}107$ cycles/byte.}
\label{fig:cost}
\end{figure}
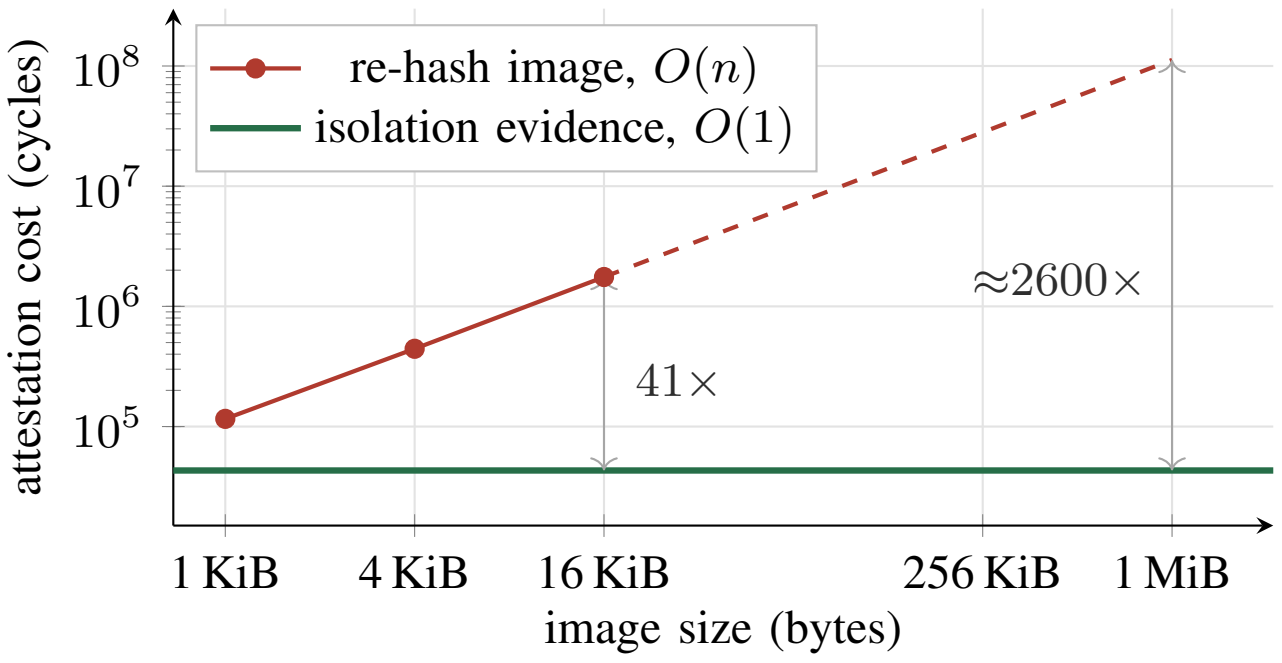

The alternative, measured on the same board, is more expensive. Re-hashing the image at attestation is retrospective, prone to \gls{toctou} races (the attested core observed the 256-byte write within two polling periods, under 5\,\textmu s, $3\,500\times$ less than the slot re-hash that would reveal it) and $O(n)$. Software \gls{sha256} runs at ${\approx}107$ cycles/byte on this core (read from \texttt{mcycle}, identical to the cycle across all 20 runs), so re-measuring 16\,KiB costs 1\,755\,080 cycles: $40.7\times$ the constant 43\,093 of the isolation evidence, and $7.6\times$ signing. 1\,MiB extrapolates to ${\approx}112$\,M cycles, ${\approx}4.5$\,s at 25\,MHz (Fig.~\ref{fig:cost}). Both checks are worth running where affordable, but only the policy evidence does not grow with the firmware it protects.

\begin{figure}[t]
\centering
\begin{tikzpicture}[x=0.64cm, y=0.34cm]
\foreach \r/\lab/\f in {%
  0/Forged report (app key)/0,
  1/Wrong active image/0,
  2/Pending update as done/2,
  3/Journal replay or splice/2,
  4/Recovery boot hidden/2,
  5/\acrshort{pmp} left unlocked/3,
  6/Journal app-writable/3,
  7/\acrshort{pmp} \acrshort{toctou}/3,
  8/\textbf{\acrshort{dma}-writable slot}/4,
  9/\textbf{\acrshort{io} \acrshort{toctou}}/4}{%
  \node[anchor=east, font=\scriptsize] at (-0.62,-\r) {\lab};
  \foreach \c in {0,...,4}{%
    \pgfmathtruncatemacro{\det}{\c<\f ? 0 : 1}%
    \pgfmathtruncatemacro{\only}{(\f>3) && (\c>3) ? 1 : 0}%
    \ifnum\det=1
      \fill[Locked, rounded corners=1pt]
        (\c-0.28,-\r-0.27) rectangle (\c+0.28,-\r+0.27);
    \else
      \fill[black!9, rounded corners=1pt]
        (\c-0.28,-\r-0.27) rectangle (\c+0.28,-\r+0.27);
    \fi
  }
}
\foreach \c/\h in {0/P0,1/P1,2/P2,3/P3,4/P4}{%
  \node[font=\scriptsize] at (\c,0.80) {\h};
}
\draw[Ink, line width=1.1pt, rounded corners=2pt]
  (3.58,-7.62) rectangle (4.42,-9.38);
\node[font=\scriptsize\bfseries, Ink, anchor=west] at (4.55,-8.5) {only P4};
\draw[line width=0.9pt, black!75]
  (-0.5,0.4) -- (-0.5,-1.5) -- (1.5,-1.5) -- (1.5,-4.5)
  -- (2.5,-4.5) -- (2.5,-7.5) -- (3.5,-7.5) -- (3.5,-9.35);
\end{tikzpicture}
\caption{Evidence-profile comparison across ten attack classes. Filled cells indicate rejection by cumulative profiles. Firmware, lifecycle, and \acrshort{cpu}-side evidence reject eight classes, while the remaining two require the update-bound \acrshort{io} policy. Table~\ref{tab:staircase} gives the hardware counterpart.}
\label{fig:coverage}
\end{figure}
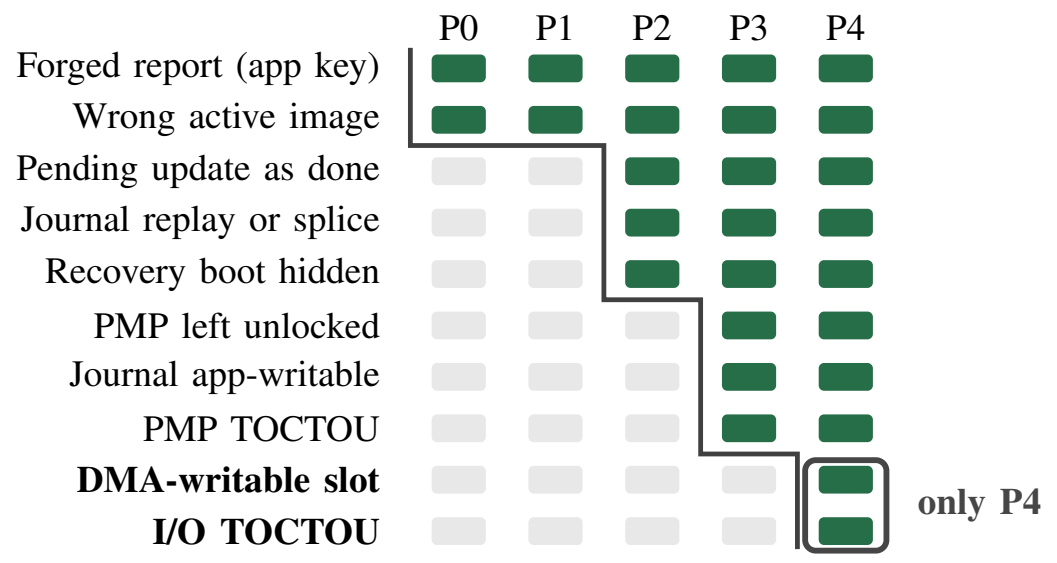

Beyond the board, a report-mutation harness extends the verifier's evaluation to ten attack classes (Fig.~\ref{fig:coverage}), each a correctly signed report of a bad state. The last two fail \emph{match} (1), the second also \emph{freshness} (2) and \emph{immutability} (3). The reset escape closed by \emph{continuity} (4) is exercised in \gls{rtl} simulation, since it requires commit state to outlive a reset, as is the latch's refusal of a second strobe. All results use a single bitstream and a single firmware image, with the policy selected at runtime, so the cases differ only in the controlled input. \gls{rtl}, firmware, verifier and scripts are available at \url{https://anonymous.4open.science/r/PULSAR-V-DATE}.

\section{Limitations}
\label{sec:limits}

The artifact mediates one master per configuration, and its \gls{iopu} table covers only the active slot: the shim, not the gate, keeps the external master off the journal. Mediation completeness rests on the topology rather than a proof, so an unmodelled port, debug master, or cache-maintenance path could bypass the gate. Reports are signed with \gls{hmac}-\gls{sha256} using a compiled-in key that any firmware can read, serving as a \gls{rom}-only asymmetric key (Sec.~\ref{sec:threat}), so the artifact is not a deployable attester. Measurement integrity is likewise assumed with all firmware in \gls{mmode}, and running the application in \gls{umode} under \gls{pmp} or Dorami-style separation~\cite{dorami} would discharge it. Only the
\gls{io} gate latches its commit state: the \gls{cpu} gate's commit digest, the image digest, and lifecycle state stay software-reported, tamper-evident only once the \gls{rom} anchor authenticates the whole commit record. The reset counter survives core resets, but not power cycles, so a deployable one requires non-volatile backing, and the reset escape is simulated only. Legitimate policy changes are indistinguishable from attacks and require re-enrolment. Finally, the core reads the slot without executing it, so the demonstration covers an undetected modification rather than its execution.

\section{Conclusion}

Post-update attestation that reports firmware identity, lifecycle, and \gls{cpu}-side isolation can certify devices whose committed image has already been corrupted by a master that the attested processor does not control. We demonstrated the attack on hardware with a \gls{dma} engine as the adversary and closed it on our prototype by binding the update commit to snapshots of both gates, the \gls{io} gate latching its own. The result is not a faster integrity check but a different verdict: an intact device is rejected when its own policy left the image writable.

\bibliographystyle{IEEEtran}
\bibliography{references}

\end{document}

%% file: acronyms.tex
% Acronyms, loaded from the preamble of main.tex.
% Use \gls{key} / \glspl{key} in running text (first use expands),
% \acrshort{key} / \acrshortpl{key} in floats, keywords and the title.
% Prior-work system names (ASSURED, PURE, VRASED, ...), standards bodies,
% companies and part numbers are names, not acronyms, and are not listed.

% --- processor and memory protection ---------------------------------------
\newacronym{cpu}{CPU}{central processing unit}
\newacronym[longplural={systems-on-chip}]{soc}{SoC}{system-on-chip}
\newacronym{mcu}{MCU}{microcontroller unit}
\newacronym{io}{I/O}{input/output}
\newacronym{pmp}{PMP}{physical memory protection}
\newacronym{epmp}{ePMP}{enhanced PMP}
\newacronym{iopmp}{IOPMP}{I/O physical memory protection}
\newacronym{iopu}{IOPU}{I/O protection unit}
\newacronym{mpu}{MPU}{memory protection unit}
\newacronym{iommu}{IOMMU}{I/O memory management unit}
\newacronym{smmu}{SMMU}{system memory management unit}
\newacronym{dma}{DMA}{direct memory access}
\newacronym{rlb}{RLB}{Rule Lock Bypass}
\newacronym{mmode}{M-mode}{machine mode}
\newacronym{umode}{U-mode}{user mode}
\newacronym{tor}{TOR}{top-of-range}
\newacronym{napot}{NAPOT}{natural alignment power-of-two}
\newacronym{na4}{NA4}{natural alignment four-byte}
\newacronym{rom}{ROM}{read-only memory}

% --- buses and peripherals --------------------------------------------------
\newacronym{axi}{AXI}{Advanced eXtensible Interface}
\newacronym{ahb}{AHB}{Advanced High-performance Bus}
\newacronym{obi}{OBI}{Open Bus Interface}
\newacronym{uart}{UART}{universal asynchronous receiver-transmitter}

% --- FPGA implementation ----------------------------------------------------
\newacronym{fpga}{FPGA}{field-programmable gate array}
\newacronym{ps}{PS}{processing system}
\newacronym{rtl}{RTL}{register-transfer level}
\newacronym{lut}{LUT}{look-up table}
\newacronym{ff}{FF}{flip-flop}
\newacronym{bram}{BRAM}{block random-access memory}
\newacronym{dsp}{DSP}{digital signal processing}

% --- security and firmware --------------------------------------------------
\newacronym{toctou}{TOCTOU}{time-of-check to time-of-use}
\newacronym{hmac}{HMAC}{hash-based message authentication code}
\newacronym{sha256}{SHA-256}{256-bit Secure Hash Algorithm}
\newacronym{ota}{OTA}{over-the-air}
\newacronym{tpm}{TPM}{Trusted Platform Module}
\newacronym{ima}{IMA}{Integrity Measurement Architecture}
\newacronym{tdisp}{TDISP}{Trusted Execution Environment Device Interface Security Protocol}
\newacronym{cca}{CCA}{Confidential Compute Architecture}